\documentstyle[amsfonts,aps,epsfig]{revtex}

\def\uth{u_{\mbox{\scriptsize th}}}
\def\ueq{u_{\mbox{\scriptsize eq}}}
\def\r{{\mbox{\bf r}}}

\begin{document}

\title{Geometric approach to the dynamic glass transition}

\author{Tom\'as S.~Grigera$^\ddagger$, Andrea Cavagna$^\star$, Irene
Giardina$^\dagger$, and Giorgio Parisi$^\ddagger$}

\address{$^\ddagger$ Dipartimento di Fisica, Unit\`a INFM and Sezione 
INFN Universit\`a di Roma ``La Sapienza'', 00185 Roma, Italy}
\address{$^\star$ Department of Physics and Astronomy, The University,
Manchester, M13 9PL, United Kingdom}
\address{$^\dagger$ Service de
Physique Th\`eorique, CEA Saclay, 91191 Gif-sur-Yvette, France}
             
\date{October 26, 2001}

\wideabs{
\maketitle

\begin{abstract}
We numerically study the potential energy landscape of a fragile
glassy system and find that the dynamic crossover corresponding to the
glass transition is actually the effect of an underlying geometric
transition caused by the vanishing of the instability index of saddle
points of the potential energy.  Furthermore, we show that the
potential energy barriers connecting local glassy minima increase with
decreasing energy of the minima, and we relate this behaviour to the
fragility of the system.  Finally, we analyze the real space structure
of activated processes by studying the distribution of particle
displacements for local minima connected by simple saddles.
\end{abstract}

}

Despite a large number of investigations, there is still much to
understand about the dynamic glass transition in supercooled liquids.
The basic problem is that, strictly speaking, there is no dynamic
transition at all. In systems known as {\it fragile} liquids
\cite{angell}, experiment finds a sharp rise of the viscosity in a
very narrow interval of temperature upon cooling. The shear relaxation
time increases by several orders of magnitude in a few degrees, and it
becomes impossible to perform an equilibrium experiment.
Nevertheless, sharp as this behaviour may be, it is not a genuine
dynamic singularity. At the other extreme of the experimental spectrum
we find {\it strong} liquids \cite{angell}, which experience a gentle
increase of the relaxation time, often according to the Arrhenius
law. Even in such systems though, when the viscosity becomes too
large, equilibrium can no longer be achieved in experimental times.

The glass transition temperature $T_g$ is conventionally defined as
that where the value of the viscosity is $10^{13}$ poise. Below $T_g$
equilibrium experiments become really hard to perform and a sample can
be considered to be in its glass phase. However, $T_g$ is just a
conventional experimental temperature, defined out of the need to mark
the onset of glassy dynamics. The attempt to give a theoretical
description of such an ill-defined ``transition'' may therefore seem
pointless.

On the one hand, this conclusion is correct for the strongest liquids:
here nothing peculiar happens close to $T_g$, and the glass transition
fully displays its purely conventional nature.  On the other hand, the
most fragile systems resist such an objection, simply by virtue of the
{\it extremely} steep increase of relaxation time in a small interval
of temperature around $T_g$.  This fact suggests that some kind of new
physical mechanism is indeed responsible for the onset of the glassy
phase in fragile supercooled liquids.  We share this view, and the aim
of this Letter is to shed some light on the nature of this mechanism.

The key idea is that the sharp dynamic crossover observed in fragile
liquids is a consequence of an underlying topological transition,
controlled by energy, rather than temperature.  More precisely, the
existence of an energy level where the instability index of the
stationary points of the potential energy vanishes is responsible for
a change in the dominant mechanism of diffusion.  If energy barriers
are large, this change in the mechanism of diffusion causes the fast
increase of the relaxation time.

This study is part of a more general program aimed at explaining
glassy dynamics in terms of properties of the potential energy
landscape.  Our method generalizes the ideas of Goldstein \cite{gold},
and Stillinger-Weber \cite{sw}, by extending to unstable stationary
points the analysis formerly restricted to minima of the potential
energy. The first steps in this direction have been done in
\cite{bradipo}, building on the ideas of \cite{laloux,selle}, and more
concrete results have been recently obtained in
\cite{ruocco,sad}. Further inspiration came from the approach of Keyes
and coworkers \cite{inm}, which related diffusion to the stability
properties of instantaneous configurations.  In the present work we
firmly establish the connection between topological properties of the
landscape and fragile glassy dynamics. Furthermore, we study the role
of potential energy barriers and we analyze the real space structure
of activated processes.

We consider a soft-sphere binary mixture \cite{SS}, a fragile model
glass-former. In addition to capturing the essential features of
fragile glasses \cite{SS,RoBaHa,SWMC,parisi}, this model can be
thermalized below $T_g$ with the efficient MC algorithm of
\cite{SWMC}.  Furthermore, previous investigations of the saddle
points have focused on Lennard-Jones systems, so it is useful to look
at a broader class of models.  Most of our data are obtained for
$N=70$ particles, but we tested our key results for $N=140$ as well.
We impose periodic boundary conditions in $d=3$ dimensions. Particles
are of unit mass and belong to one of two species $\alpha=1,2$,
present in equal amounts and interacting via a potential
\begin{equation}
{\cal V} = \sum_{i<j}^N V_{ij}(|\mbox{\bf r}_i-\mbox{\bf r}_j|) =
 \sum_{i<j}^N \left[ \sigma_{\alpha(i)} + \sigma_{\alpha(j)} \over
 |\mbox{\bf r}_i - \mbox{\bf r}_j | \right]^{12}.
\end{equation}
The radii $\sigma_\alpha$ are fixed by $\sigma_2/\sigma_1=1.2$ and
setting the effective diameter to unity, that is $(2\sigma_1)^3 +
2(\sigma_1+\sigma_2)^3+(2\sigma_2)^3 = 4 l_0^3$, where $l_0$ is the
unit of length. The density is $\rho=N/V= 1$ in units of $l_0^{-3}$,
and we set Boltzmann's constant $k_B=1$.  We obtain equilibrium
configurations at several temperatures by the swap Monte Carlo
algorithm of \cite{SWMC}. A long-range cut-off at $r_c=\sqrt{3}$ is
imposed. However, to find the stationary points we need a potential
with a continuous second derivative.  Thus, instead of simply shifting
the pair potential by a constant $C_{ij}$ (so that $V_{ij}(r \ge
r_c)=0$), we use a smooth cut-off, setting $V_{ij}(r)=B_{ij} (a-r)^3$
for $r_c < r < a$ and $V_{ij}(r)=0$ for $r \ge a$, fixing $a$,
$B_{ij}$ and $C_{ij}$ by imposing continuity.

\begin{figure}
\begin{center}
\leavevmode
\epsfxsize=\columnwidth
\epsffile{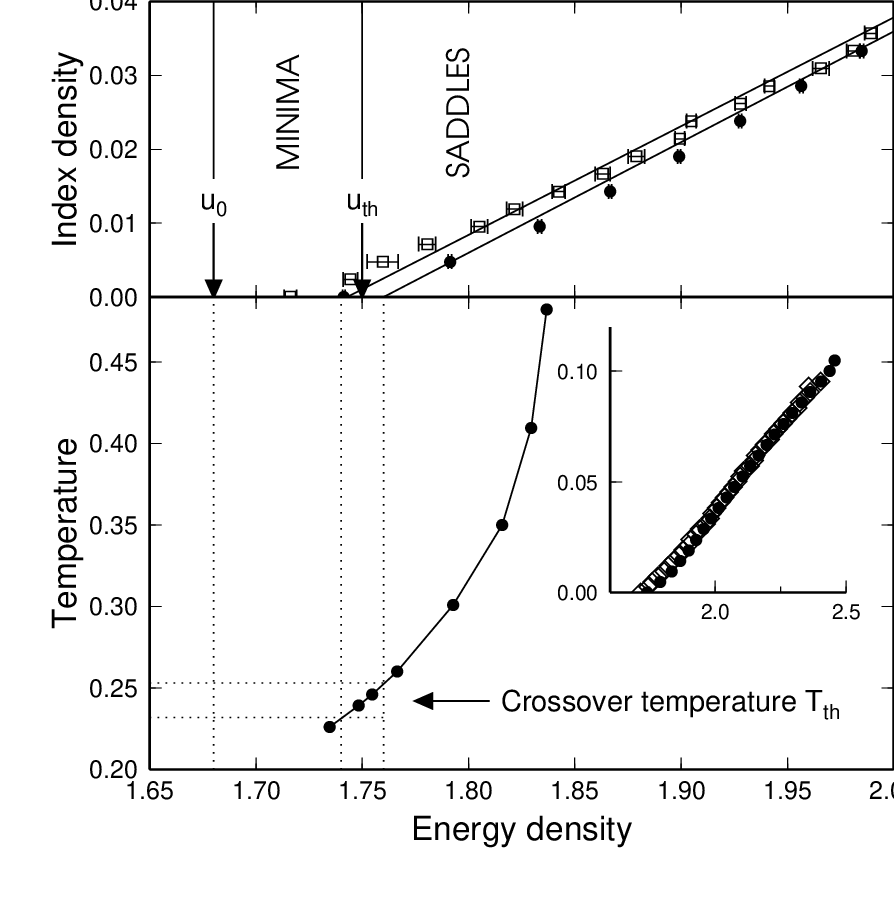}
\caption{Top: Average instability index density $k$ vs.\ potential
energy density $u$ of the stationary points.  Bottom: Temperature vs.\
equilibrium bare potential energy density, $u_b=\ueq -3/2 \, k_B T$.
Inset: $k(u)$ on the whole range sampled. Open symbols: $N=140$,
filled symbols: $N=70$.}
\label{Kappa}
\end{center}
\end{figure}

We sample the stationary points of the potential energy by quenching
the equilibrium MC configurations onto saddle points. This is done by
numerically solving the $3N$ nonlinear equations $\partial {\cal
V}/\partial \mbox{\boldmath $r$}_i = 0$ by means of a backtracking
Newton method with finite-difference approximation to the Jacobian
\cite{NR}. Once a saddle point is found, we measure its potential
energy $U$ and its instability index $K$, that is the number of
negative eigenvalues of the Hessian matrix at the saddle.

In Fig.~\ref{Kappa} (top) we plot the average index density $k=K/3N$
as a function of the potential energy density $u=U/N$. As in
\cite{sad}, we find a well defined function $k(u)$ which vanishes at a
threshold value of the energy, $\uth$. For $N=70$, a comparison
between the linear fit of the data and the last point of the curve
gives $\uth=1.75 \pm 0.01$.  For $N=140$, despite worse statistics, we
find $\uth=1.73 \pm 0.01$, consistently with the result for $N=70$.
Note that the threshold energy is not the ground state of the system,
and in fact we found minima down to $u_0=1.68$ ($N=70$).  The
threshold energy marks the border between unstable saddle points,
dominating the landscape above $\uth$, and stable minima, dominant for
$u_0< u < \uth$.  Therefore, a topological transition takes place at
$\uth$, where the stability properties of the landscape change.

A system confined to the minima-dominated region of the landscape, $u
< \uth$, must resort to barrier hopping to diffuse in phase space. It
is therefore essential to find the temperature $T_{\rm th}$ below
which this confinement takes place.  To this end we must realize that
a system trapped in a single potential well has a potential energy
density equal to the bare energy of the bottom of the well, plus a
vibrational contribution proportional to $k_B T$. Therefore, it is the
{\it bare} potential energy of the system which we must compare with
the threshold energy \cite{sad}. We can write the bare energy as
$u_b(T)=\ueq (T) - 3/2 \, k_B T$.  When the bare energy $u_b(T)$ drops
below the threshold $\uth$, the system is effectively confined to the
minima dominated region of the landscape. For $N=70$ this happens at
$T_{\rm th} = 0.242 \pm 0.012$ (Fig.~\ref{Kappa}, bottom). Note that,
unlike previous investigations \cite{ruocco,sad}, for $N=70$ we
thermalize the system {\it below} the threshold, giving an accurate
determination of $T_{\rm th}$ \cite{shame}.  Hence, the dynamic effect
of the topological transition at $\uth$ must be a qualitative change
in the mechanism of diffusion at $T_{\rm th}$.

The fact that barrier crossing becomes the main mechanism of diffusion
below $T_{\rm th}$ does not necessarily imply a slowing down of the
dynamics: large energy barriers at $\uth$ are also needed,
i.e. $\Delta U(\uth)$ substantially larger than $k_B T_{\rm th}$.  The
value of $\Delta U(\uth)$ can be estimated as the average difference
in energy between simple saddles ($K=1$) and threshold minima
($K=0$). This difference can be extracted from the slope of $k(u)$, as
$\Delta U(\uth)\approx 1/[3\,k'(\uth)]$ \cite{laloux,bradipo}. Note
that the slope of $k(u)$ for $N=70$ and $N=140$ is the same. This is
consistent with the fact that activated processes are local in space
(as we will discuss later) and therefore barriers do not depend on the
size of the system. We find $\Delta U(\uth)\approx 2.2 \approx 10\,
k_B T_{\rm th}$.  This is an important result: at the temperature
$T_{\rm th}$ where activation becomes dominant, potential energy
barriers are already very large compared to the available thermal
energy.  Activation is therefore highly inefficient at the temperature
where for the first time it is actually needed.  This, we believe, is
the most striking feature of very fragile liquids and it confirms the
conjecture of \cite{bradipo} that the fragility of a system is higher
the larger the potential energy barriers at $T_{\rm th}$. We thus
predict that a sharp slowing down of the dynamics must occur at
$T_{\rm th}$.  Note that the change in the mechanism of diffusion at
$T_{\rm th}$ {\it would not} be accompanied by a dynamic slowing down
if barriers at $T_{\rm th}$ were small. In this case there would
rather be a fragile-to-strong crossover at $T_{\rm th}$, as discussed
in \cite{bradipo}.

To test our prediction about the slowing down we must find the
temperature marking the onset of glassiness.  For soft-spheres, this
is generally accepted to be $T_c\approx 0.226$ \cite{RoBaHa}. This
value is affected by the same arbitrariness as the experimental $T_g$,
since an arbitrary time scale (set by the simulation times) is
involved \cite{RoBaHa}.  However, as stressed in the introduction, the
slowing down of fragile glasses is so sharp that it makes sense to
define a $T_c$, as long as one keeps the above proviso in mind.
\begin{figure}
\begin{center}
\leavevmode \epsfxsize=\columnwidth \epsffile{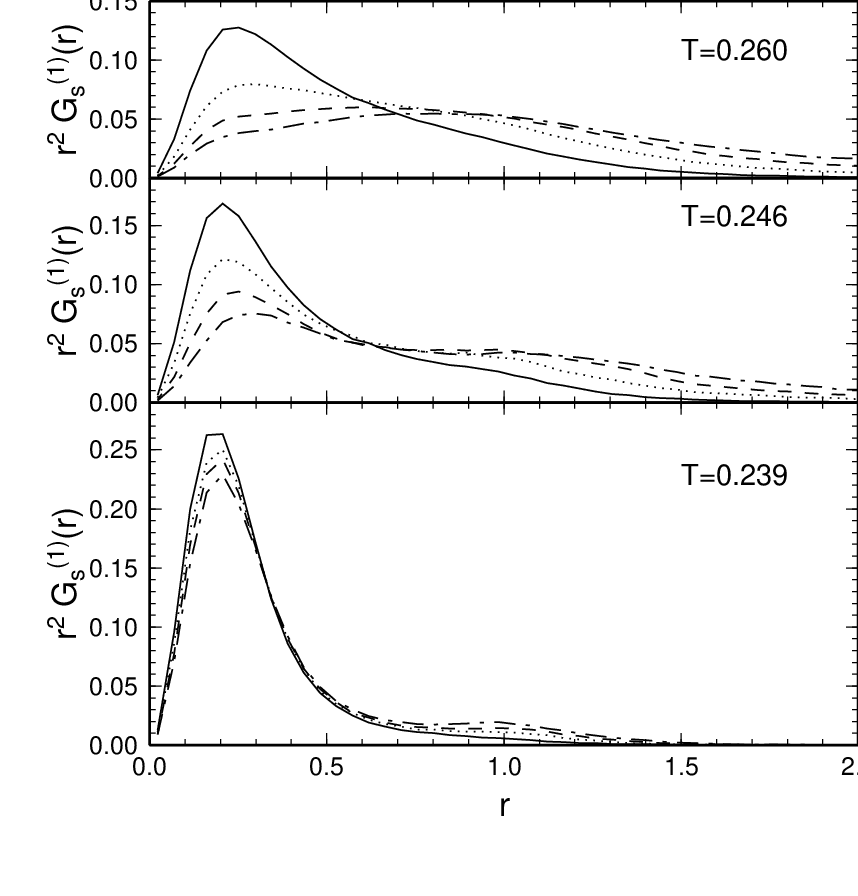}
\caption{Van Hove self correlation functions for particle type 1, at
times $t = 88$ (full line), 177 (dotted), 265 (dashed) and 353
(dash-dotted). $N=70$.}
\label{vanHove}
\end{center}
\end{figure}
Given that we use a non-standard cut-off for the potential, we perform
an independent determination of $T_c$ for $N=70$.  To this end we
compute the van Hove self-correlation function $G_s^\alpha(r,t)$ from
configurations sampled in a molecular dynamics (MD) run which uses
equilibrium MC configurations as starting points. Of course, the MD
simulation falls out of equilibrium at higher temperatures than the MC
swap dynamics.  $G_s^\alpha$ is defined as
\begin{equation}
G_s^{(\alpha)} (\r,t) = {1 \over N_\alpha} \sum_{i=1}^{N_\alpha}
   \left\langle \delta[ \r_i(t) - \r_i(0) - \r ] \right\rangle \ .
\end{equation}
The probability that a particle of type $\alpha$ has moved a distance
$r$ in a time $t$ is proportional to $r^2 G_s^{(\alpha)}(r,t)$, which
is plotted in Fig.~\ref{vanHove} for several times and temperatures.

A reliable dynamic diagnostic for $T_c$ is to look at the evolution of
the first peak of $r^2 G_s^{(\alpha)}(r,t)$ \cite{RoBaHa}.  In the
liquid phase, the peak moves to the right and rapidly becomes Gaussian
(top panel of Fig.~\ref{vanHove}). On the other hand, in the glassy
phase it takes a huge time to reach the hydrodynamic limit, and the
simulations show an unmoving peak whose area very slowly decreases as
a secondary peak grows (bottom panel). The middle panel shows an
intermediate situation. On this basis, we estimate $T_c \approx 0.24$,
not far from the standard $T_c$.  This value is consistent with the
topological transition temperature $T_{\rm th} \approx 0.242$ we found
above, a result which strongly supports our scenario.  Let us stress
the difference between $T_c$ and $T_{\rm th}$: The first depends on
the time scale of the experiment and it can be sensibly defined only
if the dynamic crossover is sharp.  The latter marks the point where
activation starts ruling the dynamics, and it is uniquely defined.  In
fact, $T_{\rm th}$ has the same nature as the critical temperature of
mode coupling theory \cite{mct}.

Our estimate of the barriers can be criticized, since the {\it
average} distance in energy between minima and simple saddles neglects
the requirement that they must be connected to each other. To test our
estimate we perform for $N=70$ a direct sampling of the potential
energy barriers. Starting from a simple saddle we follow the gradient
in the two opposite directions along the unstable eigenvector,
obtaining two connected minima.
\begin{figure}
\begin{center}
\leavevmode
\epsfxsize=\columnwidth
\epsffile{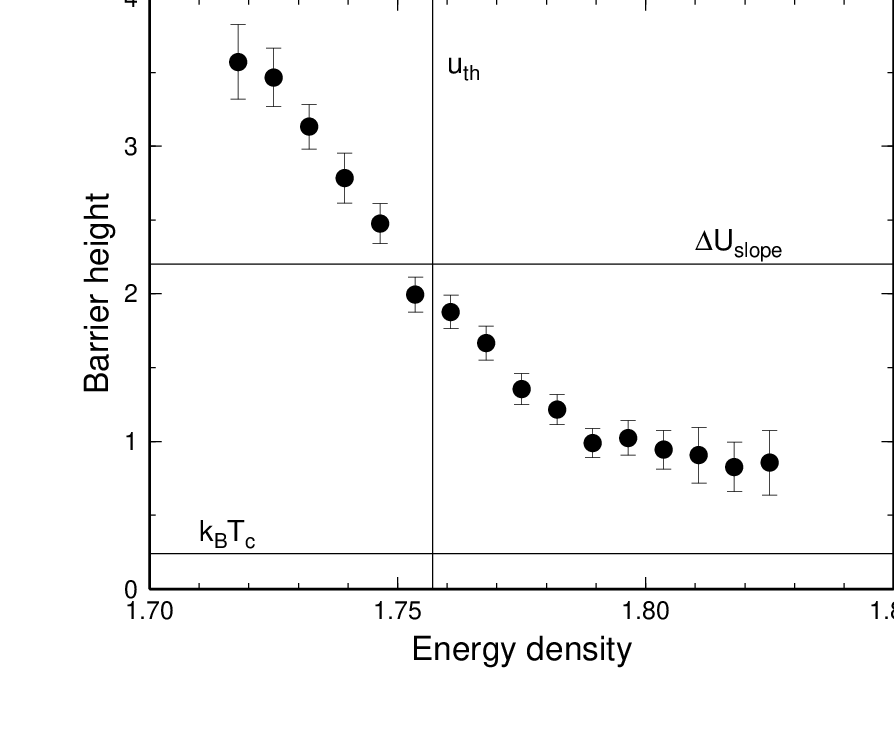}
\caption{Average potential energy barriers as a function of the
potential energy density of the adjacent minimum. Points are an
average over $1652$ barriers. $N=70$.}
\label{Barrier}
\end{center}
\end{figure}
In Fig.~\ref{Barrier} we plot the average barrier size $\Delta U$ as a
function of the energy density $u$ of the adjacent minimum.  On the
same plot we report the value of $\uth$ ($N=70$), and the estimate of
$\Delta U(\uth)$ obtained from the slope of $k(u)$: this estimate
agrees with the value that can be read off from the plot.  We conclude
that the function $k(u)$ provides the threshold energy {\it and} the
potential energy barriers at the threshold.

A second important piece of information is contained in
Fig.~\ref{Barrier}: the typical barriers grow with decreasing energy.
The consequence is that below $T_{\rm th}$ the dynamic slowing down is
enhanced not only by the decrease of the thermal energy available for
activation, but also by the increase of the typical barriers.  Thus,
below $T_{\rm th}$, we expect super-Arrhenius behaviour of the
relaxation time \cite{dan}.

A brief comment on barrier crossing is in order here.  Thermal
activation must be introduced within a canonical description, since in
the micro-canonical ensemble the total energy is conserved.  In other
words, an activated transition is performed by a sub-system, with the
rest acting as a thermal bath. The sub-system must be much smaller
than the total system, and indeed activated processes involve a finite
number of particles.  For this reason, potential energy barriers
associated with such processes are finite in the thermodynamic limit,
implying that simple saddles and minima have the same potential energy
density for $N\to\infty$.  Yet, the equilibrium potential energy
density is of order $k_B T$ above minima.  This fact does {\it not}
imply that barriers are easy to overcome, {\it nor} that thermal
activation is irrelevant, but simply that activated processes involve
a finite number of particles.
\begin{figure}
\begin{center}
\leavevmode \epsfxsize=\columnwidth \epsffile{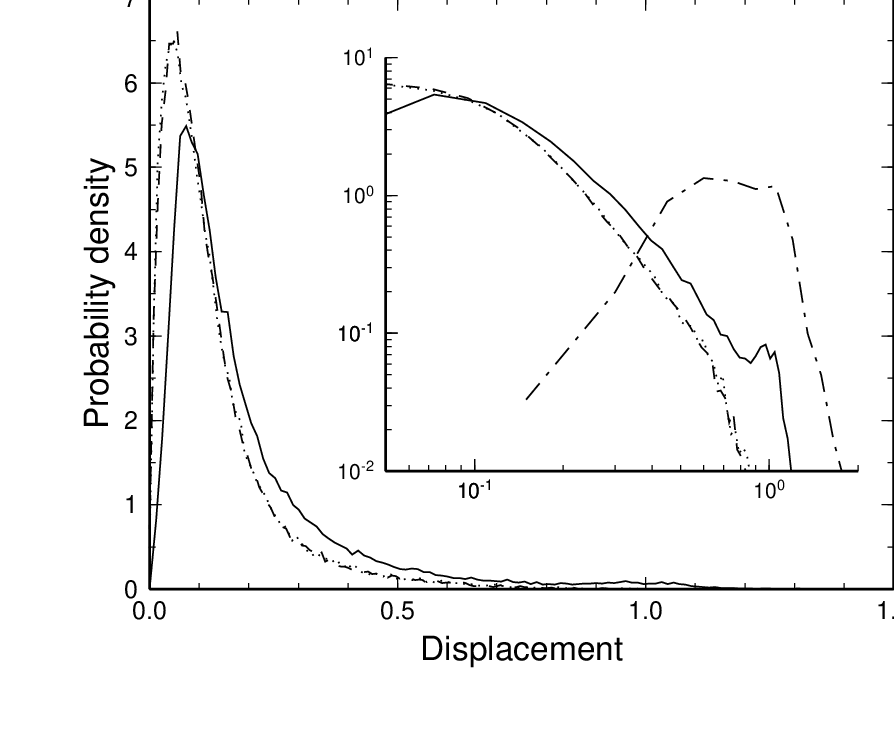}
\caption{Distribution of particle displacements. Full line: distance
between the two minima.  Dashed (dotted) line: distance between the
saddle and the right (left) minimum.  Inset: same plot in a log-log
scale. Dashed-dotted line: distribution of largest displacement.
$N=70$.}
\label{displacement}
\end{center}
\end{figure}
As we have shown, activated processes are crucial below $T_{\rm th}$
and therefore a real-space description of barrier crossing is very
important.  To this end we computed the distribution of the {\it
displacement}, that is the distance between the position of a particle
in a minimum and its position in the crossing-connected minimum
(Fig.~\ref{displacement}) \cite{gio}.  To interpret this result we
need first to fix a reference distance: Fig.~\ref{vanHove} shows that,
even in the glassy phase, particles can easily travel a distance $r_m
\approx 0.5$. The primary peak in the displacement distribution
indicates that the large majority of particles moves less than $r_m$,
while a small secondary peak can be seen at $r\approx 1$, involving
only $\approx 2$ particles \cite{glo}.  From the radial distribution
function (not shown) we know that $r\approx 1$ is the nearest-neighbor
distance for type 1 particles. These facts thus suggest that in this
system activated processes involve a small number of nearest neighbor
particles exchanging positions, while many particles move a small
amount to make way for them.  We also measure the {\it largest}
displacement for each pair of minima and find that its distribution
has no secondary peak at short distances, confirming the above
interpretation.  Finally, the distribution of the displacements
between minimum and intermediate saddle shows no secondary peak at
large distances, consistently with the fact that on the saddle the
exchanging particles have not completed their transition yet.

In this Letter we argued that glassy slowing down in fragile liquids
is caused by the presence of a topological transition.  Potential
energy barriers are much larger than the available thermal energy at
the transition, and they increase with decreasing energy. Activated
processes involve small numbers of particles, each moving a distance
of the order of the nearest-neighbor separation.

AC was sup\-por\-ted by EPSRC-GR/L97698, TSG partly by CO\-NI\-CET
(Ar\-gen\-ti\-na).  We thank J.L. Barrat, J.-P. Bouchaud, B. Doliwa,
G. Ehrhardt, J.P. Garrahan, A. Heuer, E. Marinari, V. Martin-Mayor,
M.A. Moore, F. Ricci-Tersenghi, A. Stephenson and F. Thalmann.


\begin{references}
\bibitem{shame} For $N=140$ we could not equilibrate below $\uth$, and
we had not enough points to perform an extrapolation of
$u_b(T)$. However, it has been shown that standard static properties
(as the bare energy) do not show significant finite-size effects for
$N>60$ in soft spheres \cite{parisi} and Lennard-Jones systems
\cite{heuer}. Thus, $u_b^{[N=70]}$ can be compared with $u_{\rm
th}^{[N=140]}$, giving $T_{\rm th}^{[N=140]} \approx 0.22$.
\bibitem{dan} The distribution of $\Delta U$ at fixed $u$ is fairly
broad, with a small fraction of barriers much smaller than $k_B T_{\rm
th}$. These rare, but fast barriers may be connected to non-activated
off-equilibrium dynamics (AC thanks D. Kivelson and G. Tarjus for
important discussions on this point).
\end{references}
\end{document}